\documentclass[aps,showpacs,preprint,superscriptaddress,nofootinbib,11pt]{revtex4-1}
\usepackage{amsmath}
\usepackage{bm}
\usepackage[english]{babel}
\usepackage{subfigure}
\usepackage{graphicx}
\usepackage{color}

\begin{document}
\title{On the contribution of exchange interactions to the Vlasov equation}

\author{J. Zamanian}
\email[E-mail address: ]{jens.zamanian@physics.umu.se}
\affiliation{Department of Physics, Ume{\aa} University, SE-901 87 Ume{\aa},
Sweden}

\author{M. Marklund}
\email[E-mail address: ]{mattias.marklund@chalmers.se}
\affiliation{Department of Physics, Ume{\aa} University, SE-901 87 Ume{\aa},
Sweden}
\affiliation{Department of Applied Physics, Division for Condensed Matter
Theory, Chalmers University of Technology, SE-412 96 G\"oteborg, Sweden}

\author{G. Brodin}
\email[E-mail address: ]{gert.brodin@physics.umu.se}
\affiliation{Department of Physics, Ume{\aa} University, SE-901 87 Ume{\aa},
Sweden}

\begin{abstract}
Exchange effects play an important role in determining the equilibrium
properties of dense matter systems, as well as for magnetic phenomena. There
exists an extensive literature concerning, e.g., the effects of exchange
interactions on the equation of state of dense matter. Here, a
generalization of the Vlasov equation to include exchange effects is
presented allowing for electromagnetic mean fields, thus incorporating some
of the dynamic effects due to the exchange interactions. Treating the
exchange term perturbatively, the correction to classical Langmuir waves in
plasmas is found, and the results are compared with previous work. It is
noted that the relative importance of exchange effects scales similarly with
density and temperature as particle dispersive effects, but that the overall
magnitude is sensitive to the details of the specific problem. The
implications of our results are discussed.
\end{abstract}

\maketitle

\section{Introduction}

Recently there has been an increased interest in the properties of quantum
plasmas \cite{Haas-book,manfredi2006,Shukla-Eliasson-RMP}. The interest has
been stimulated by applications to for example quantum wells \cite%
{Manfredi-quantum-well}, spintronics \cite{Spintronics}, plasmonics \cite%
{Atwater-Plasmonics}, and laser plasma interaction on solid density targets 
\cite{glenzer-redmer}. Historically, there is a vast literature concerning
effects from the quantum regime on the statistical equilibrium properties of
matter systems (see e.g. \cite{Eliezer-etal} and references therein for a
discussion). Concerning dynamical problems, much of the work within the
framework of kinetic theory has been based on the celebrated Wigner-Moyal
approach \cite{Haas-book,manfredi2006,Shukla-Eliasson-RMP}. While the
corresponding Wigner-Moyal equation takes particle dispersive effects from
first principles into account, as well as being compatible with Fermi-Dirac
statistic, there are still several types of quantum effects that are not
included in this model. This includes e.g. spin dynamics \cite%
{Zamanian-2010-NJP}, various types of relaxation processes \cite%
{Kadanoff-Baym} and exchange effects \cite%
{nacht83,brosnens84,Zamanian-exchange}.

In this work we will ignore magnetization dynamics associated with the
electron spin \cite{Zamanian-2010-NJP}, relaxation processes associated with
particle correlations \cite{Kadanoff-Baym}, and concentrate on the dynamical
effects due to exchange interactions. This will of course limit the model in
terms of applicability, but it will serve to highlight the particular nature
of the exchange interactions. For long scale lengths the Wigner-Moyal
reduces to the Vlasov equation. However, the relative importance of exchange
effects does not generally diminish with increasing scale lengths \cite%
{Zamanian-exchange}, and for dense plasmas exchange terms may provide
important corrections to the Vlasov equation. Generalizing the treatment of
Ref. \cite{Zamanian-exchange} to include electromagnetic mean fields,
summing over the spin states, we derive an evolution equation for the
distribution function in the Hartree-Fock approximation, applicable in the
regime of long scale lengths. Dropping the exchange term the Vlasov equation
is recovered. The theory is illustrated by considering Langmuir waves in a
plasma with the temperature much larger than the Fermi temperature. The
exchange term is treated as a small perturbation and the correction to the
linear Langmuir dispersion relation is calculated. By comparing with similar
results for low-frequency ion-acoustic waves we note that the relative
importance of exchange effects scales in the same way with temperature and
density for the two cases. However, while the scaling is the same for high-
and low-frequency waves, the overall magnitude in the exchange effects
differs quite strongly. The implications of our results are pointed out, and
the relation to other theories of quantum plasmas, in particular density
functional theory \cite{TDFT,Manfredi-DFT}, is discussed

%%%%%%%%%%%%%%%%%%%%%%%%%%%%%%%%%%%%%%%%%%%%

\section{Exchange effects in plasmas: The electromagnetic case}

%%%%%%%%%%%%%%%%%%%%%%%%%%%%%%%%%%%%%%%%%%%%
In a previous paper, Ref.\ \cite{Zamanian-exchange}, we considered the long
scale-length limit of quantum exchange effects in electrostatic plasmas of
fermions. This was done by first deriving the evolution equation for the
Wigner distribution in the Hartree-Fock approximation where the
anti-symmetric part to the mean field electric potential was included. The
equation was then simplified using the assumption that the thermal de
Broglie wavelength of the particles is much smaller than the typical scale
length. In this approximation the part of the evolution equation which is
independent of exchange effects simplifies from an integro-differential
equation, i.e. the Wigner-Moyal equation, to a Vlasov type of equation. For
the exchange terms some of the involving integrals are then solvable. We
further considered the case of a completely spin unpolarized plasma which
further simplified the exchange terms.

In Ref.\ \cite{Zamanian-exchange} only the electric field interaction was
considered. Here we make the extension to also include a magnetic field in
the formalism, although the particle motion is still assumed to be
non-relativistic. We focus on the long scale length limit, i.e. consider a
characteristic scale length $L$ fulfilling $L\gg \hbar /mv_{T}$, where $%
v_{T} $ is the thermal velocity, $\hbar =h/2\pi $ where $h$ is Planck's
constant and $m$ is the mass. In this regime, after lengthy algebra, the
evolution equation of the Wigner distribution $f(\mathbf{x,p,}t)$ \cite%
{wigner32} is found to be 
\begin{widetext} 
\begin{align} 
	\partial_t f (\mathbf x, \mathbf p, t) & 
	+ \frac{\mathbf p}{m} \cdot \nabla_x f(\mathbf x, \mathbf p, t) 
	+ q \left[ \mathbf E (\mathbf x,t) + \frac{\mathbf p}{m} \times \mathbf B (\mathbf x,t) \right]
		\cdot \nabla_p f(\mathbf x ,\mathbf p, t) 
	\notag \\ & 
	= 
	- \frac{1}{2} \partial_p^i \int d^3\! r \, d^3\! p' \,\, e^{ - i \mathbf r \cdot \mathbf p' / \hbar} 
	[\partial_{r}^i V (- \mathbf r)] 
	f \left( \mathbf x + \frac{\mathbf r}{2} , \mathbf p + \frac{\mathbf p'}{2}, t \right) 
	f \left( \mathbf x + \frac{\mathbf r}{2} , \mathbf p - \frac{\mathbf p'}{2}, t \right)  
	\notag \\ &
	- \frac{i \hbar}{8}  
	\partial_p^i \partial_p^j \cdot \int d^3\! r \, d^3\! p' \,\, 
	e^{ - i \mathbf r \cdot \mathbf p' / \hbar} 
	[\partial_{r}^i V (- \mathbf r) ]
\notag \\ &
	\quad \times 
	\left[ f \left( \mathbf x + \frac{\mathbf r}{2} , \mathbf p - \frac{\mathbf p'}{2},  t \right) 
	\left( \overleftarrow \partial_x^j 
	- \overleftarrow \partial_p^k q \left[ \partial_j A_k ( \mathbf x ) \right]
	- \overrightarrow \partial_x^j 
	+ q \left[ \partial_x^j A_k ( \mathbf x ) \right] \overrightarrow \partial_p^k \right)
	f \left( \mathbf x + \frac{\mathbf r}{2} , \mathbf p + \frac{\mathbf p'}{2} , t \right)  \right]
\label{evolution}  
\end{align} 
\end{widetext}where $V(\mathbf{r})=q^{2}/(4\pi \epsilon _{0}r)$, $q$ is the
charge, $\epsilon _{0}$ is the vacuum permittivity, $\mathbf{p}$ and $%
\mathbf{p}^{\prime }$ denote kinetical momentum which is related to the
canonical momentum $\mathbf{p}_{c}$ according to $\mathbf{p}=\mathbf{p}_{c}-q%
\mathbf{A}$ where $\mathbf{A}$ is the vector potential. Furthermore, $%
\mathbf{x}$ and $\ \mathbf{r}$ denotes position vectors, and we have also
defined $\partial _{r}^{i}=\partial /\partial r_{i}$ and an arrow above a
differential operator indicates in which direction it acts. In the terms of
the right hand side sums over repeated indices are understood, with $%
i,j,k=1,2,3$. Note that the equation still explicitly contains the vector
potential and is not gauge invariant in its current form. However, the
Wigner function still gives gauge independent results for observables such
as the charge and current densities, which are given by $\rho _{c}=q\int
d^{3}pf$ and $\mathbf{j=(}q/m)\int d^{3}pf\mathbf{p}$ respectively, see
Ref.\ \cite{Zamanian-2010-NJP} for further discussion.

It is of interest to compare Eq. (\ref{evolution}) with the Hartree-Fock
equations that are commonly used in atom physics and solid state physics.
Much of the work using the Hartree-Fock equations is limited to the
time-independent situation, whereas our Eq. (\ref{evolution}) covers the
fully time-dependent case. The time-independent equation can still be used
to calculate e.g. the influence of exchange effects on the free particle
dispersion relation in an homogenous electron gas \cite%
{Gell-man,Ashcroft-Mermin}, but the equation must be generalized to the
time-dependent case in order to cover the case of a dynamically varying mean
field. Eq. (\ref{evolution})) is obtained from a Wigner transform of the
density matrix in the time-dependent Hartree-Fock equations. For certain
types of problems the Wigner-transform is not helpful, as the number of
independent variables are increased and the general complexity seemingly
becomes higher. However, for rather broad cases of problems in plasma
physics, whenever the solution to the Vlasov equation is a reasonable first
approximation, the phase-space formalism resulting from the Wigner transform
is highly useful. The reason is that analytical first order solutions to the
Vlasov equation often can be found when the corresponding solutions for the
wave function is less than straightforward to deduce. These analytical
solutions are then a good starting point for a perturbative approach, as
demonstrated by Ref. \cite{Zamanian-exchange}, and as will be further
illustrated here.

\section{Exchange effects on Langmuir waves}

The equation above is quite complicated and in general some further
approximations are needed. In Ref.\ \cite{Zamanian-exchange} we studied the
exchange effects on the damping of ion acoustic waves. This was done by
linearizing the equation and using an iterative approach where the lowest
order (i.e. exchange independent) solution was inserted in the terms on the
right hand side. Here we study another example of electrostatic
oscillations, namely Langmuir waves \cite{note1}. To find the dispersion
relation, we linearize the evolution equation \eqref{evolution} together
with the Poisson equation 
\begin{equation}
\nabla \cdot \mathbf{E}=\frac{1}{\epsilon _{0}}\sum_{s}q_{s}\int f_{s}d^{3}p,
\label{poisson}
\end{equation}%
where the sum is over the species in the plasma. To be specific we will here
consider the the particles to be electrons, drop the species index on the
distribution function, and let $q_{e}=-e$. The positively charged ions
constitute a neutralizing background.

We start with Eq.\ \eqref{evolution} and make the ansatz 
\begin{equation}
f=f_{0}+f_{1}\exp [i(kz-\omega t)]\quad \text{and}\quad \mathbf{E}=\hat{%
\mathbf{z}}E_{1}\exp [i(kz-\omega t)],
\end{equation}%
where $\hat{\mathbf{z}}$ is a unit vector in the z-direction. Inserting this
into the linearized evolution equation \eqref{evolution} we obtain 
\begin{eqnarray}
&& 
	i \left( -\omega +\frac{kp_{z}}{m}\right) f_{1}(\mathbf{p})
	= 
	- q \mathbf{E} \cdot \nabla _{p} f_{0} (\mathbf{p})  \notag \\
&&
	- \frac{1}{2} \partial _{p}^{i} \int d^{3} p^{\prime} 
	e^{- i\left( \mathbf{p}^{\prime } - \hbar k\hat{\mathbf{z}}/2\right) \cdot \mathbf{r}/\hbar}
	[\partial _{r}^{i}V(- \mathbf{r})]
	\left[ 
		f_{1} \left( \mathbf{p}+\frac{\mathbf{p}^{\prime }}{2}\right) 
		f_{0} \left( \mathbf{p} - \frac{\mathbf{p}^{\prime }}{2}\right) 
		+ f_{0} \left( \mathbf{p} +\frac{\mathbf{p}^{\prime }}{2} \right)
		f_{1} \left( \mathbf{p} - \frac{\mathbf{p}^{\prime} }{2} \right) 
	\right]   
\notag
\\
&&
	+ \frac{\hbar k}{8}\partial _{p}^{i} \partial _{p}^{z}
	\int d^{3}p^{\prime} 
	e^{- i \left( \mathbf{p}^{\prime } - \hbar k \hat{\mathbf{z}} / 2 \right) \cdot \mathbf{r} / \hbar }
	[\partial _{r}^{i} V(- \mathbf{r})]
	\left[ f_{0}\left( \mathbf{p} - \frac{\mathbf{p}^{\prime }}{2}\right) f_{1}\left( \mathbf{p}+\frac{\mathbf{p}^{\prime }}{2}\right) 
	- f_{0} \left( \mathbf{p}+\frac{\mathbf{p}^{\prime }}{2} \right) 
	f_{1} \left( \mathbf{p}-\frac{\mathbf{p}^{\prime }}{2} \right) \right] .
\notag \\ &&
\end{eqnarray}
At this stage it is possible to solve the integrals over $\mathbf{r}$ as
this can be recognized as the Fourier transform of the derivative of the
Coulomb potential. Since we are aiming for the lowest order quantum
mechanical correction, the result is expanded to leading order in $\hbar
k/(mv_{T})$. Since we focus on the regime where the exchange terms will only
give a small correction we may insert the lowest order solution (where
exchange effects are neglected) in the integrals of the right hand side,
that is we let 
\begin{equation}
f_{1}\rightarrow f_{1}^{(0)}=\frac{qE_{z}}{i\left( \omega -kp_{z}/m\right) }%
\frac{\partial f_{0}}{\partial p_{z}}.  \label{zeroth}
\end{equation}%
Assuming that the thermodynamic temperature is much higher than the Fermi
temperature $T_{F}$ the appropriate distribution function is given by the
Maxwell-Boltzmann distribution 
\begin{equation}
f_{0}(\mathbf{p})=\frac{n_{0}}{(2\pi mk_{B}T)^{3/2}}\exp \left( -\frac{%
p_{\perp }^{2}+p_{z}^{2}}{2mk_{B}T}\right) ,  \label{maxwell-boltzmann}
\end{equation}%
where $n_{0}$ is the equilibrium density and $T$ is the temperature of the
electrons and $k_{B}$ is the Boltzmann constant. For further refinement of
the current calculation one may add semiclassical corrections to the
Maxwell-Boltzmann distribution function. After solving for $f_{1}$ to first
order in the exchange term and inserting this into the linearized version of
the Poisson equation \eqref{poisson} we get 
\begin{equation}
k-\omega _{p}^{2}k\left( \frac{1}{\omega ^{2}}+\frac{3k^{2}v_{T}^{2}}{\omega
^{4}}\right) + I=0,  \label{6}
\end{equation}%
where 
\begin{equation}
I = 
	\frac{ \hbar^2 k^2 \omega_p^4 }{2 \pi  m^2 v_T^3 } 
	\int d u_{\perp} d u_{z} d w_z 
	\frac{1}{\left( \omega -  k v_T w_z \right)^2}     
	\frac{ u_{\perp}  u_{z} }{ u_{\perp}^2 + u_{z}^2 } 
	\frac{( w_z - u_{z})( w_z + u_{z} ) }{ \left[ \omega - k v_T (w_z - u_{z})  \right] } 
	\exp \left( - u_{\perp}^2 - u_{z}^2 - w_z^2   \right)
\label{I}
\end{equation}%
where $v_{T}^{2}=k_{B}T/m$ and $\omega _{p}=(n_{0}e^{2}/m \epsilon_0)^{1/2}$ is
the plasma frequency. Since we are interested in corrections to Langmuir
waves where $\omega \approx \omega _{p}\gg kv_{T}$ and since we neglect
Landau damping, we may expand the denominator in Eq.\ \eqref{I} to the
lowest non-vanishing order in $k$. After this the integrals can be
straightforwardly solved and the resulting dispersion relation reads 
\begin{equation}
	\omega^{2} = \omega _{p}^{2} + 3 k^{2} v_{T}^{2} \left( 1 - \frac{1}{90}
	\frac{ \hbar^{2} \omega_{p}^{2} }{m^{2} v_{T}^{4} }\right) .  
\label{DR}
\end{equation}%
After correcting a slight numerical error in equation (49) of Ref.\ \cite{Roos} the result there is exactly a
factor two larger than our result above. 
The difference is due to that Ref.\ \cite{Roos} does not 
take into the spin part of the wavefunction\footnote{Note that it is the full many-body 
wavefunction that should be anti-symmetric with respect to particle interchange, not just the 
spatial part.} which over estimates the exchange correction 
by a factor of two. We note that the relative
magnitude of exchange effects scales with temperature and density as $%
H^{2}=\hbar ^{2}\omega _{p}^{2}/m^{2}v_{T}^{4}$, with the same quantum
parameter $H^{2}$ that appears in many other types of problems \cite%
{Haas-book,manfredi2006,Shukla-Eliasson-RMP}. \ 

There is also an interest to compare Eq. (\ref{DR}) with results from
density functional theory (DFT). While there is a limited number of results
to compare with, a rather commonly used expression based on the local
density approximation is 
\begin{equation}
V_{x}=\frac{0.985\kappa }{4\pi }\frac{h^{2}\omega _{p}^{2}}{mv_{F}^{2}}%
\left( \frac{n}{n_{0}}\right) ^{1/3}  \label{Exchange-pot}
\end{equation}%
for the exchange potential (see e.g. Refs. \cite%
{exchange-pot-I,exchange-pot-II,Manfredi-DFT}), deduced  assuming the
opposite ordering to our case, i.e. $T\ll T_{F}$. Here $\kappa =(3\pi
^{2})^{2/3}$ is simply a numerical factor $n$ is the number density and $%
n_{0}$ is the unperturbed number density. The same formalism also gives
contributions from particle correlations, but this should not be included in
the comparison. The potential (\ref{Exchange-pot}) could either be
substituted as and effective potential into the Schr\"{o}dinger equation
(see e.g. \cite{exchange-pot-I}) or it could be included in a quantum fluid
equation after a Madelung transformation is made \cite{Manfredi-DFT}. 

Next, in order to compare the exchange term of Eq. 10 with our own result,
we employ a fluid model. For 1-dimensional spatial variations and
electrostatic fields, the momentum equation of the fluid becomes%
\begin{equation}
\frac{\partial u}{\partial t}+u\frac{\partial u}{\partial z}=\frac{q}{m}E+%
\frac{1}{m}\frac{\partial V_{x}}{\partial z}+\frac{\hbar ^{2}}{2m^{2}}\frac{%
\partial }{\partial z}\left( \frac{\partial ^{2}(\sqrt{n})/\partial z^{2}}{%
\sqrt{n}}\right) -\frac{1}{mn}\frac{\partial P}{\partial z}
\label{Fluid-momentum}
\end{equation}%
Here $u$ is the fluid velocity, $P$ is the fluid pressure and the third term
of the right hand side is the Bohm de Broglie potential that accounts for
particle dispersive effects. Combining (\ref{Fluid-momentum}) with the
continuity equation (which is the same as in the classical case) and
Poisson's equation it is straightforward to calculate the contribution to
the Langmuir dispersion relation from the exchange potential. Since particle
dispersive effects have been neglected in our model we should neglect the
contribution from the Bohm de Broglie term for consistency, but this term
gives a contribution $\propto k^{4}$ in the dispersion relation that vanish
in the long wavelength limit in any case. The degree of agreement with the
thermal term in Eq. (\ref{DR}) depend on the equation of state for $P$,
which is not our focus here, since this issue had been previously studied by
many others. Focusing on the exchange contribution solely we deduce that if
we use (\ref{Fluid-momentum}) instead of the quantum kinetic theory our
exchange correction term is replaced according to 
\begin{equation}
\frac{1}{90}\frac{\hbar ^{2}\omega _{p}^{2}}{m^{2}v_{T}^{4}}\rightarrow 
\frac{0.985\kappa }{4\pi }\frac{\hbar ^{2}\omega _{p}^{2}}{m^{2}v_{F}^{4}}
\label{kinetic-fluid}
\end{equation}%
It should be stressed that since $T>T_{F}$ for the left hand side
expression, and the opposite ordering holds for the right hand side
expression, we cannot expect the expressions to agree in any detail. The
first difference is that $v_{T}^{4}\rightarrow v_{F}^{4}$ as expected from
the different orderings. In a very rough sense we do have a qualitative
agreement, since the scaling with plasmon energy over kinetic energy is the
same (with the kinetic energy given by $k_{B}T$ and $k_{B}T_{F}$
respectively). However, the magnitude of the terms are quite different, as $%
0.985\kappa /4\pi $ is of the order of unity, whereas our coefficient is much
smaller (in agreement with Ref. \cite{Roos}). We would like to point out that while the potential
(\ref{Exchange-pot}) is known to give good results in some restricted cases
including e.g. applications to quantum wells \cite{Manfredi-DFT,exchange-pot-I}, it has not been optimized for a case similar to the one we have been studying.

\section{Summary and discussion}

The main result of the present paper is the inclusion of exchange effects in
the Vlasov equation (see Eq. (\ref{evolution})), where the derivation of 
\cite{Zamanian-exchange} is generalized to include electromagnetic mean
fields. \ In our treatment we have neglected collisional effects
(correlations), whose relative importance roughly scale as $N^{-1}$, where $%
N=n\lambda _{D}^{3}$ is the number of particles in a Debye cube. This should
be compared to the relative importance of exchange effects, that tend to
scale as $uH^{2}$, where $u$ is a dimensionless factor that is dependent on
the details of the specific problem at hand and $H^{2}=\hbar ^{2}\omega
_{p}^{2}/m^{2}v_{T}^{4}$. Clearly $H^{2}$ and $N$ increase with density and
decrease with thermal energy, although the two scalings are not identical,
and the opposing regimes $H^{2}>N$ and $H^{2}<N$ are both possible. As long
as both exchange effects and collisions are small enough to be treated
perturbatively, however, their respective contributions can be computed
separately and added afterwards. Naturally a qualitative difference between
collisions and exchange effects is that the former increase entropy and
drives the system towards the equilibrium distribution.

Let is exemplify our discussion of parameters with a concrete example. In
addition to solid state plasmas exchange effects can be of significance for
plasmas in the inertial confinement fusion (ICF) regime. Here the electron
density may reach values of the order $n\simeq 10^{32}\mathrm{m}^{-3}$,
which give us $T_{F}\simeq 7 \times 10^{6}K.$ In ICF experiments the
temperature may vary significantly during the different stages. Before the
heating phase the temperature correspond to a partially degenerate plasma ($%
T\sim T_{F}$) and at the latter stages the plasma temperature can reach $%
T\gg T_{F}$ \cite{ICF-book}. To pick an example where our perturbative
treatment is valid but when the exchange effects still can be significant,
we may let $H^{2}\simeq 0.1$ which correspond to the case where $T$ is a
factor $3$ larger than the Fermi temperature (such that a Maxwell-Boltzmann
distribution is a reasonable first approximation), i.e. $T\simeq 2 \times
10^{7}K$.

A difficulty when trying to estimate the importance of exchange effect is
that the value of the factor $u$ is hard to predict. In the present problem
it turned out that $u\approx 0.011$, as can be seem from Eq. (\ref{DR}).
Following the details of the calculation leading up to (\ref{DR}) we see
that this modest value does not come from a small parameter of any sort, but
is just due to successive multiplications of a few numbers each slightly
smaller than unity. A consequence is that the exchange effects on linear
Langmuir waves is most likely too small to be detected under such
conditions. However, if the problem of Langmuir waves is replaced by
ion-acoustic waves, the situation happens to be different \cite%
{Zamanian-exchange}. The relative importance of exchange effects still has
the overall scaling of $H^{2}$, but the value of $u$ is much larger. In Ref. 
\cite{Zamanian-exchange} both the correction to the real and the imaginary
value to the ion-acoustic dispersion relation was calculated, and the
corresponding values of $u$ came out as $\mathrm{Re}$ $u\approx 0.80$ and $%
\mathrm{Im}$ $u\approx 3.0$.

Thus our current results indicate that exchange phenomena is negligible for
fast phenomena (on the plasmon time-scale), whereas previous works \cite%
{Zamanian-exchange} supports that exchange effects give highly significant
corrections for low-frequency phenomena (occurring on the ion-acoustic
time-scale). As the understanding of the numerical factor $u$ is fairly
limited, it should be stressed that this generalization may be far to bold,
however. Nevertheless the difference in $u$ for various phenomena can have
important theoretical consequences. A common approach when modelling quantum
plasma kinetically is to use the Wigner-Moyal equation \cite%
{Haas-book,manfredi2006,Shukla-Eliasson-RMP} which includes particle
dispersive effects but ignores exchange effects. When applying the
Wigner-Moyal equation to collective phenomena (where the self-consistent
field is important), the relative importance of the particle dispersive
effects also scales as $H^{2}$ (provided the scale lengths are sufficiently
short, of the order of the thermal de Broglie wavelength). Thus for Langmuir
waves it may be safe to keep particle dispersive effects and drop exchange
effects, as $u\ll 1$ assures the validity of this approach. For ion-acoustic
waves, on the other hand, the same approach is questionable. The relative
magnitude of particle dispersive effects still scales as $H^{2}$ (for short
enough scale-lengths), but as the magnitude of $u$ is no longer smaller than
unity exchange effects is equally important. A reservation that must be made
here is that the results above applies for $T_{F}\ll T$, whereas many works
have studied the opposite ordering. Qualitatively one would expect that much
of the results described above remains the same, but with $H^{2}\rightarrow
\hbar ^{2}\omega _{p}^{2}/m_{e}^{2}v_{F}^{2}$ when $T\ll T_{F}$. To some
degree this assertion is supported by comparison with the results presented
in Ref. \cite{Manfredi-DFT}. Here DFT-calculations lead to fluid equations
with corrections due to the exchange effect of the order $\hbar ^{2}\omega
_{p}^{2}/m_{e}^{2}v_{F}^{2}$. \ An interesting prospect for future work is
to make more detailed comparisons of the theories presented here with
calculation based on time-dependent density functional theory \cite{TDFT}.


\begin{thebibliography}{99}
\bibitem{note1} We hence postpone the studies of magnetized plasmas and
electromagnetic waves even though our new model, Eq.\ \eqref{evolution},
allows for this.
\end{thebibliography}
\end{document}